\documentstyle [12pt]{article}
\newcommand{\beq}{\begin{equation}}

\newcommand{\eeq}{\end{equation}}

\newcommand{\beqa}{\begin{eqnarray}}
\newcommand{\eeqa}{\end{eqnarray}}
\begin{document}
\baselineskip=1.6\baselineskip
\hbox{\hskip 4.0 true in CCUTH-93-01}
\vbox{\vskip 0.5 true in}

\centerline{\large \bf Stability Analysis of Sum Rules }
\centerline{\large \bf for pion Compton Scattering}
\vbox {\vskip 0.3 true cm}
%
\centerline{Claudio Corian\`{o}
\footnote{Stiftelsen Blanceflor Boncompagni-Ludovisi Fellow}
and Hsiang-nan Li $^{2}$}
\vskip 0.3cm
\centerline{ ${^1}$ Institute for Theoretical Physics, University of
Stockholm,}
\centerline{Box 6730, 113 85 Stockholm, Sweden}
\centerline {and}
\centerline{Department of Radiation Sciences, Uppsala University}
\centerline{Box 535, S-75121, Uppsala, Sweden}
\centerline{$^2$ Department of Physics, National Chung-Cheng University,}
\centerline{Chia-Yi, Taiwan, R.O.C.}

\vbox {\vskip 0.5 true in}
\centerline{ABSTRACT}


\newpage
Recently it has been suggested \cite{CRS} that QCD sum rule analysis can
be extended from their usual setting, which includes two and
three-point functions \cite{SVZ,NR,IS},
to processes of Compton type at moderate values
of the Mandelstam invariants $s$, $t$ and $u$.
Sum rules for a specific combination of the two invariant amplitudes
in pion Compton scattering
corresponding to different helicities, $H_1 +H_2$, have been given
in ref.~\cite{CRS}, while the
evaluation of relevant power corrections has been discussed in \cite{cla}.
Comparison of sum rule predictions - in the local duality limit -
with predictions from a modified perturbative calculation which
includes Sudakov suppression
for soft gluon exchange \cite{LS} has been made in \cite{CL}.
This new approach to Compton scattering, complementary to the usual
perturbative one, is
interesting in order to investigate the transition from
non-perturbative to perturbative QCD in these processes.
It also suggests that the non-perturbative information on this type
of reactions can be parameterized by the lowest dimensional vacuum
condensates through the operator product expansion of four interpolating
currents.

For the simplest Compton reaction $\pi +\gamma \to \pi + \gamma$
two steps are still missing in completing the above programme.
They are:  1) a stability analysis of the sum rule, and
2) the derivation of an individual sum rule for
each of the two invariant amplitudes, $H_1$ and $H_2$.
While here our attention is focused on the first point, we reserve the
discussion of the second point to a future work.
A stability analysis of the sum rule proposed in \cite{cla} is crucial
in order to give physical justification to previous works
and to allow further extension of the new formalism to many other
similar reactions. As stated in \cite{CL}, the angular dependence of
Compton scattering, absent in form factors, allows us to compare
perturbative QCD and sum rule predictions in a nontrivial way.
Our aim in this letter is first to characterize the
stability region of sum rules for pion Compton scattering, and,
second, to compare the sum rule predictions with
the perturbative QCD approach based on the modified factorization
formula.

Before proceeding to study the stability of the complete sum rule,
we recall the conclusions of
\cite{CRS,cla}, and quote the relevant results below.
A sum rule for the $sum$ of the two helicities
of pion Compton scattering has been derived by studying the
following correlation function of local currents \cite{CRS}
\begin{eqnarray}
\Gamma_{\sigma\mu\nu\lambda}(p_1^2,p_2^2,s,t)&=&i\int{\rm d}^4x\,{\rm d}^4y
\,{\rm d}^4z\exp
(-ip_1\cdot x+ip_2\cdot y-iq_1\cdot z)
\nonumber \\
& &\times \langle 0|T\left(\eta_{\sigma}(y)J_{\mu}(z)J_{\nu}(0)
\eta_{\lambda}^{\dagger}(x)\right)|0\rangle \; ,
\label{tp}
\end{eqnarray}
where
\begin{eqnarray}
J_{\mu}=\frac{2}{3}\bar{u}\gamma_{\mu}u -\frac{1}{3}\bar{d}\gamma_{\mu}d,
\;\;\;\;\;\;
\eta_{\sigma}=
\bar{u}\gamma_5\gamma_{\alpha}d
\label{jd}
\end{eqnarray}
are the electromagnetic and axial currents respectively of up and down quarks.
These currents interpolate with the two invariant amplitudes of the
scattering process.
The two photons carry on-shell momenta $q_1$ and $q_2$, and are physically
polarized. The momenta of the two pions are denoted as $p_1$ and $p_2$,
with $s_1=p_1^2$ and $s_2=p_2^2$ there virtualities. We also define
$s=(p_1 +q_1)^2$, $t=(p_2-p_1)^2$ and $u=(p_2-q_1)^2$ for the Mandelstam
invariants, which obey the relation $s + t +u=s_1 + s_2$.
The invariant amplitudes are obtained
selecting a specific time ordering in eq.~(\ref{tp}) and projecting the
two axial currents onto single-pion states. $H_1$ and $H_2$
are isolated from the matrix element
\beq
M_{\mu\nu}= i\int d^4z \, e^{-iq_1\cdot z}
\langle p_2|T\left(J_{\mu}(z)J_{\nu}(0)\right) |p_1 \rangle
\label{mnula}
\eeq
by the expansion
\beq
M_{\mu\nu}= H_1(s,t) e^{(1)}_{\mu}e^{(1)}_{\nu} +
H_2(s,t) e^{(2)}_{\mu}e^{(2)}_{\nu},
\label{h1h2}
\eeq
where $e^{(1)}$ and $e^{(2)}$ are helicity vectors defined in \cite{CRS,cla}.

A sum rule relates the timelike region  of $s_1$ and $s_2$, where the
resonant contribution to $M_{\mu\nu}$ is located,
to the
so called ``deep Euclidean region", where an operator product expansion
(OPE) for the four-current correlator is made, through
a dispersion relation.
In the timelike region the spectral
density appearing in the dispersion relation is commonly modeled by
\begin{eqnarray}
\Delta_{\sigma\mu\nu\lambda}&=&f_{\pi}^2p_{1\lambda}p_{2\sigma}(2\pi)^2
\delta(p_1^2)\delta(p_2^2)M_{\mu\nu}\nonumber \\
& &+\Delta^{\rm pert}\left[1-\theta(s_0-p_1^2)\theta(s_0-p_2^2)\right]\;,
\label{m}
\end{eqnarray}
where the second term, the continuum contribution, which is
nonvanishing only for $p_1^2$, $p_2^2 > s_0$, is chosen as the
perturbative spectral density $\Delta^{\rm pert}$,
the leading term of OPE in the deep Euclidean region.
Notice that the OPE of the spectral density includes,
besides the perturbative part,
non-perturbative power corrections proportional to the condensates of
quarks and gluons, which are determined in close analogy with the
canonical approach developed for the form factor case \cite{NR,IS}.

A Borel transform then acts on both regions in order
to enhance the resonant contribution respect to the continuum.
Given the fact that the dispersion relation for Compton scattering
involves only a finite domain of the complex $p_i^2$ plane,
a modified version of Borel transform \cite{CRS}
\begin{equation}
{\cal B} = \int_{C}\frac{dp_1^2}{M_1^2}\int_{C}\frac{dp_2^2}{M_2^2}
e^{-p_1^2/M_1^2}e^{-p_2^2/M_2^2}
\left (1 - e^{-(\lambda^2-p_1^2)/M_1^2} \right ) \left (1 - e^{-(\lambda^2-
p_2^2)/M_2^2} \right)
\label{bo}
\end{equation}
is introduced,
with $C$ a contour of radius $\lambda^2$, which is kept finite
in order to exclude the $u$-channel resonances from
the phenomenological ansatz for the spectral density.
In fact, $\lambda^2$ can vary from $s_0$ to $(s +t)/2$, though it was set
approximately to the value $(s+t)/4$ in \cite{CRS,cla}.
This modification introduces an extra unphysical parameter $\lambda^2$,
in addition to the usual Borel mass $M^2$,
and therefore, the stability of the sum rule has to be found in the
two-dimensional $M^2$-$\lambda^2$ plane.
The factor
$1-\exp(-(\lambda^2-p_i^2)/M^2)$ in (\ref{bo}),
which is new compared to the standard Borel transform
\cite{NR,IS}, is to ensure that the transform is still regular
when $C$ crosses the branch cut in the $p_i^2$ plane.

The resulting sum rule for $H=H_1+H_2$ can be expanded asymptotically for the
large invariant $Q^2$ \cite{cla},
\beqa
Q^2={1\over 4}\left(s_1 + s_2 -t + \sqrt{(s_1 +s_2 -t )^2 - 4 s_1 s_2}
\right)\;,
\label{q}
\eeqa
as
\beqa
&&{f_\pi}^2H(s,t)\left({s(s+t)\over -t}\right)
\left(1- e^{-\lambda/M^2}\right)^2=\nonumber \\
& &\hspace*{0.5cm}\left(\int_{0}^{s_0}ds_1\int_{0}^{s_0}ds_2
\rho^{\rm pert}+\frac{\alpha_s}{\pi}\langle G^2 \rangle
\int_{0}^{\lambda^2}ds_1\int_{0}^{\lambda^2}ds_2\rho^{\rm gluon}\right)
\nonumber \\
& &\hspace*{0.5cm}\times
e^{-(s_1+s_2)/M^2}\left(1- e^{-(\lambda^2-s_1)/M^2}\right)
\left(1- e^{-(\lambda^2-s_2)/M^2}\right)\nonumber \\
& &\hspace*{0.5cm}+C^{\rm quark}\pi\alpha_s \langle
(\bar{\psi}\psi)^2 \rangle
\left(1 - e^{-\lambda^2/M^2} \right)\;,
\label{s1}
\eeqa
where the perturbative, gluonic and quark contributions are
given by, respectively,
\beqa
\rho^{\rm pert}&=&{ 2560 Q^{14} \tau^{\rm pert}(s,Q^2,s_1,s_2)\over
3\pi^2  (s-2 Q^2 )(4 Q^4- s_1 s_2)^5 (s_1 s_2-2 Q^2 s)}\;,\nonumber \\
\rho^{\rm gluon}&=&{ 20480 Q^{22}(s - 2 Q^2)
\tau^{\rm gluon}(s,Q^2,s_1,s_2)\over
27 s (2 Q^2 -s_1)^2  (2 Q^2 -s_2)^2 (4 Q^4 - s_1 s_2)^5 (2 Q^2 s - s_1 s_2)
(4 Q^4 -2 Q^2 s + s_1 s_2)^2}\;, \nonumber \\
C^{\rm quark}&=&-{16\over 9}
{(8 M^2 s +4 s^2 +2 M^2 t +4 s t + t^2)\over M^4 t}\;,
\label{oc}
\eeqa
with
\beqa
\tau^{\rm pert}&=&(s- Q^2)^2
(2 Q^4 s s_1 - Q^2 s^2 s_1 + 2 Q^4 s s_2 - Q^2 s^2 s_2 \nonumber \\
&&- 2 Q^4 s_1 s_2 - 6 Q^2 s s_1 s_2 + 3 s^2 s_1 s_2)\;, \nonumber \\
\tau^{\rm gluon}&=&-8 Q^{12} s -8 Q^{10} s^2 + 68 Q^8 s^3 -64 Q^6 s^4+
16 Q^4 s^5 \nonumber \\
& &+8 Q^{10} s s_1 +8 Q^8 s^2 s_1 -108 Q^6 s^3 s_1 +104 Q^4 s^4 s_1
-26 Q^2 s^5 s_1 \nonumber \\
& &+8 Q^{10} s s_2 +8 Q^8 s^2 s_2 -108 Q^6 s^3 s_2
+ 104 Q^4 s^4 s_2 -26 Q^2 s^5 s_2\;. \nonumber \\
& &
\label{gluonic}
\eeqa
Note that the perturbative contribution comes only from the interval
$(0,s_0)$, since the contribution from $(s_0,\lambda^2)$ is cancelled
by that from the phenomenological side of the sum rule.
The gluon and quark condensates,
$\langle G^2\rangle$ and $\langle (\bar{\psi}\psi)^2\rangle$,
take the values
\begin{eqnarray}
&&\frac{\alpha_s}{\pi}\langle G^2\rangle=1.2\times 10^{-2}{\rm GeV}^4\nonumber
\\
&&\alpha_s\langle (\bar{\psi}\psi)^2\rangle=1.8\times 10^{-4}{\rm GeV}^6\;.
\label{vev}
\end{eqnarray}

The approximation made in the calculation of the gluonic power correction,
with the gluonic coefficient obtained by integrating
its double discontinuity along the real interval
$(0,\lambda^2)$ for each of $s_1$ and $s_2$, has been discussed in
\cite{cla}. A similar spectral representation of this coefficient has been
used in \cite{IS} in the investigation of the form factor sum rule.
In this latter case the integration interval of such spectral
representation is $(0,\infty)$,
since the spectral density for the triangle diagram
is regular for all $s_1$, $s_2 > 0$ and $t<0$
(see refs.~\cite{IS} and \cite{cla} for details).

Based on the formulas (\ref{q})-(\ref{vev}),
we perform the stability analysis of eq.~(\ref{s1}).
The $\lambda^2$ dependence of $H$ in a wide range of the Borel mass
$M^2=4$-8 GeV$^2$ at
$s=20$, $|t|=4$ and $s_0=0.7$ GeV$^2$ is shown in fig.~1. Obviously,
all the curves located in the region marked by the vertical bars, in which
the power corrections do not exceed 50\% of the perturbative contribution,
increase rapidly with $\lambda^2$. The power corrections always
dominate for $M^2$ below 4 GeV$^2$. This result indicates that there is
not a stable region for $H$ when the radius $\lambda^2$ of the
Borel transform is varied. A similar behavior is observed
for other choices of $s$, $t$ and $s_0$.
As already mentioned above, such dependence is not present in the
form factor case mainly because the variables $s_i$ in
the spectral representation for the coefficient of the gluonic
power correction runs up to infinity \cite{IS}.

We do not expect the strong $\lambda$ dependence
in the sum rule for two reasons:
1) $\lambda$ is an unphysical parameter, and a physical quantity
like the invariant amplitude $H$
should be insensitive to it; 2) the $\lambda$ dependence,
introduced by the Borel transform, should cancel from both sides of
the sum rule. Therefore, the sensitivity of our results to the radius
of the Borel transform just implies that
the phenomenological model in (\ref{m})
is too simple to maintain a stability of the
sum rule in the $M^2$-$\lambda^2$ plane.

We propose here a modified phenomenological model in order to remove
this spurious dependence:
\begin{eqnarray}
\Delta_{\sigma\mu\nu\lambda}&=&f_{\pi}^2p_{1\lambda}p_{2\sigma}(2\pi)^2
\delta(p_1^2)\delta(p_2^2)M_{\mu\nu}\nonumber \\
& &+\Delta^{\rm OPE}\left[1-\theta(s_0-p_1^2)\theta(s_0-p_2^2)\right]\;,
\label{nm}
\end{eqnarray}
with the continuum contribution replaced by $\Delta^{\rm OPE}$,
which is the same as the full spectral expression on the OPE
side of the sum rule.
This modification makes sense, because the region with large virtualities
$p_1^2$, $p_2^2 > s_0$ can be regarded as perturbative, and an OPE
is allowed. With this choice we are requiring that
the hadronic spectral density for the continuum
from $s_0$ to $\lambda^2$ truncates not only
the perturbative part (the lowest order contribution), but also
the power corrections, of the OPE side of the sum rule.
Since the contributions from the region $(s_0, \lambda^2)$ have been removed
by the above cancellation, $s_i$'s  never reach the upper bound $\lambda^2$,
and the remaining $\lambda$ dependent factors
$1- \exp[-(\lambda^2-s_i)/M^2]$ from the transform (\ref{bo})
can be dropped. Therefore, the overall
dependence on the radius of the Borel transform
disappear completely from the sum rule.

Using eq.~(\ref{nm}), (\ref{s1}) is modified to
\beqa
&&{f_\pi}^2H(s,t)\left({s(s+t)\over -t}\right)=\nonumber \\
& &\hspace*{0.5cm}\left(\int_{0}^{s_0}ds_1\int_{0}^{s_0}ds_2
\rho^{\rm pert}+\frac{\alpha_s}{\pi}\langle G^2 \rangle
\int_{0}^{s_0}ds_1\int_{0}^{s_0}ds_2\rho^{\rm gluon}\right)
e^{-(s_1+s_2)/M^2} \nonumber \\
& &\hspace*{0.5cm}+{\hat C}^{\rm quark}\pi\alpha_s
\langle (\bar{\psi}\psi)^2 \rangle\;,
\label{s2}
\eeqa
with the modified quark contribution
\beqa
{\hat C}^{\rm quark}=-{16\over 9}
{(8 M^4 s^2 +8 M^4 s t +8 M^2 s^2 t +8 M^2 s t^2 +4 s^2 t^2
+4 s t^3+ t^4)\over M^4 t^3}\;.
\label{nc}
\eeqa
Note the change of the upper bound from $\lambda^2$ to $s_0$
in the integral for the gluonic power correction, which is
due to the cancellation from the phenomenological side.

Before studying the new sum rule (\ref{s2}), we shall examine
how the modified parametrization for the continuum affects the sum rule
calculation of the pion form factor.
This is a significant check for the process we aim to discuss,
since it has been shown \cite{CRS} that
Compton scattering has a strong similarity
to the corresponding form factor case at moderate $s$ and $t$
and at a fixed angle.
The modified sum rule for pion form factor $F_{\pi}$
can be derived easily based on \cite{IS}:
\beqa
&&\frac{f_\pi^2}{4}F_{\pi}(Q^2)=\nonumber \\
&&\hspace*{0.5cm}\left(\int_{0}^{s_0}ds_1\int_{0}^{s_0}ds_2
\rho_{\pi}^{\rm pert}+\frac{\alpha_s}{\pi}\langle G^2 \rangle
\int_{0}^{s_0}ds_1\int_{0}^{s_0}ds_2\rho_{\pi}^{\rm gluon}\right)
e^{-(s_1+s_2)/M^2} \nonumber \\
& &\hspace*{0.5cm}+C_{\pi}^{\rm quark}\pi\alpha_s
\langle (\bar{\psi}\psi)^2 \rangle\;,
\label{s3}
\eeqa
with
\beqa
\rho_{\pi}^{\rm pert}&=&\frac{3Q^4}{16\pi^2}\frac{1}{\delta^{7/2}}
\left[3\delta(s_1+s_2+Q^2)(s_1+s_2+2Q^2)-\delta^2-5Q^2(s_1+s_2+Q^2)\right]\;,
\nonumber \\
\rho_{\pi}^{\rm gluon}&=&\frac{1}{48M^4}\left[
\frac{-Q^2(s_1+s_2)+(s_1-s_2)^2-2\delta-2Q^4}
{\delta^{3/2}}\right.\nonumber \\
& &\left.+\frac{4Q^2(s_1+s_2+Q^2)}{\delta}
+\frac{s_1+s_2+Q^2}{M^2\delta^{1/2}}\right]\;,\nonumber \\
C_{\pi}^{\rm quark}&=&\frac{52}{81M^4}\left(1+\frac{2Q^2}{13M^2}\right)\;,
\nonumber \\
\delta &=&(s_1 +s_2 -t)^2 - 4 s_1 s_2 \;.
\label{p1}
\eeqa
Note that the expression for the quark power correction
$C_{\pi}^{\rm quark}$ based on the modified phenomenological model is the
same as in \cite{NR,IS}. A Borel
transform picks up only the residue of the pole $1/p_1^2p_2^2$ in the quark
power correction, and there is not such a pole term in the integral
from $s_0$ to $\lambda^2$ on the phenomenological
side of the sum rule. Therefore,
the quark part in $\Delta^{\rm OPE}$ does not contribute,
when the Borel transform is applied. There is not cancellation from
the phenomenological side for the quark power correction.
The difference of the modified expression (\ref{nc}) in the case of
pion Compton scattering from the corresponding one in
(\ref{oc}) is due to the neglect of the suppressing factors
mentioned above, which results in the change of the spectral density.

Results for $F_{\pi}$ obtained from the stability analysis of (\ref{s3})
are shown in fig.~2
with the best values of $s_0=0.7$ and $M^2=2$ GeV$^2$ substituted into
the sum rule. It is evident that the behavior of $F_{\pi}$ in $Q^2$
is indeed in good agreement with experimental data \cite{B}
even after these amendments, and consistent with
those derived in \cite{NR,IS}, which are based on the standard
phenomenological model (\ref{m}).

Now we proceed to the stability analysis of (\ref{s2}) following a
method similar to \cite{NR,IS}. Since the $\lambda$ dependence has been
removed completely, we concentrate simply on the variation of $H$ with
respect to $M^2$.
The $M^2$ dependence of $H$ for $s_0=0.5-0.7$
GeV$^2$ at $s=20$ and $|t|=4$ GeV$^2$ is displayed in fig.~3.
The region on the right-hand side of vertical bars is the one where
the power corrections do not exceed 50\% of the perturbative contribution.
It is obvious that as $s_0=0.6$ GeV$^2$
there is the largest $M^2$ interval,
in which $H$ is approximately constant. Therefore, $s_0=0.6$ GeV$^2$
is the best choice which makes both sides of the sum rule most coincident.
This value of the duality interval is close to that given
in the form factor case, and consistent with its conjectured value
0.7 GeV$^2$ in \cite{CL}.
Different sets of $s$ and $t$ have been investigated. The best value of
$s_0$ does not vary significantly, and $H$ is almost constant
within the range $2 < M^2 < 6$ GeV$^2$.

Results for $H$ at different scattering angles of the photon, $\theta$,
$\sin(\theta/2)=-t/s$, with $s_0=0.6$ and $M^2=4$
GeV$^2$ are exhibited in fig.~4, where $|H|$ denotes the magnitude of
$H$. Basically, they show a similar dependence on
angles and momentum transfers $|t|$ to those derived using local duality
approximation \cite{CL}, but with the magnitude lower
by 20\% only at small $|t|$. These predictions are
compared to the perturbative predictions obtained from the modified
factorization formula \cite{CL}. The transition to perturbative QCD at
about $|t|=4$ GeV$^2$ and $-t/s=0.5$ $(\theta=40^o)$,
where the perturbative contributions begin to dominate, is observed.
Sum rule results are always smaller for $-t/s=0.6$ $(\theta=50^o)$,
and always larger for $-t/s=0.2$ $(\theta=15^o)$, than the
perturbative results.
This is also consistent with the conclusion in \cite{CL}.

We have analyzed in more detail the sum rule which describes the
behaviour of the sum of the two helicities
of pion Compton scattering close to the resonant region,
and its dependence on the two parameters, $M^2$ and $\lambda^2$,
which characterize the modified Borel transform.
We have seen that the strong dependence of the sum rule on the radius of
the transform can be removed consistently under the assumption
that resonances from mass higher than the duality interval $s_0$
contribute equally both to the
phenomenological side and to the OPE side.
The results for the two helicities are found to
be stable under the new phenomenological parametrization for the continuum
in a wide variation of the Borel mass, and is characterized by a
local duality interval
which takes a value similar to the form factor case.
A more detailed discussion of these issues
on individual sum rules for the two helicities
will be considered elsewhere \cite{CL1}.

\vskip 0.5cm
We thank Prof. G. Sterman for helpful discussions. The work of H.N.L.
was supported by the National Science Council of ROC under Grant No.
NSC82-0112-C001-017. C.C. thanks the Physics Department of Lund University
for hospitality, and the Physics Institute
of Academia Sinica for hospitality and for partial financial support.

\newpage
\centerline{\large \bf Figure Captions}
\vskip 1.0cm
\noindent
{\bf Fig. 1} Dependence of $H$ on $\lambda^2$ at $s=20$, $|t|=4$ and
$s_0=0.7$ GeV$^2$ for
$M^2=4$ (solid line), 6 (dashed line), and 8 GeV$^2$ (dotted line).
\vskip 0.3cm
\noindent
{\bf Fig. 2} Dependence of $F_{\pi}$ on $Q^2$ derived from the modified
phenomenological model (\ref{nm}) with $s_0=0.7$ and $M^2=2$ GeV$^2$
(solid line). Results from
ref.~\cite{IS} (dashed line) and from ref.~\cite{NR} (dotted line), and
experimental data (dots) are also shown.
\vskip 0.3cm
\noindent
{\bf Fig. 3} Dependence of $H$ on $M^2$ at $s=20$ and $|t|=4$ GeV$^2$
for (a) $s_0=0.7$ GeV$^2$, (b)
$s_0=0.6$ GeV$^2$, and (c) $s_0=0.5$ GeV$^2$.
\vskip 0.3cm

\noindent
{\bf Fig. 4} Dependence of $|t||H|$ on $|t|$ derived from the full analysis
of QCD sum rules with $s_0=0.6$ and $M^2=4$ GeV$^2$
(solid lines) for (a) $-t/s=0.6$ $(\theta=50^o)$,
(b) $-t/s=0.5$ $(\theta=40^o)$, and (c) $-t/s=0.2$ $(\theta=15^o)$.
Corresponding results from the modified perturbative QCD
calculation (dashed lines) are also shown.
\end{document}